# A LITTLE KNOWN ASTRONOMER IN THE LATE ISLAMIC PERIOD. A STUDY OF QÂSIM [c]AlÎ AL-QÂYINÎ'S MANUSCRIPTS

**Marjan AKBARI**
Department of Physics, Tohoku University, 980-8578 Sendai, Japan
E-mail: makbari@sspp.phys.tohoku.ac.jp

**Abstract:** It is widely believed that the advance of science in the Islamic world after the mid-fifteenth century A.D. suffered a decline. For the purposes of examining this belief, a manuscript by Qâsim [c]Alī al-Qâyinî (ca. A.D.1685) was chosen based on previous works which considered it a valuable source on the history of optics that had not been studied before. After studying his major optical manuscript, titled *Manâẓir wa Marâyâ*, it was found very interesting that the majority of al-Qâyinî's propositions relating to natural phenomena were not merely geometrical definitions, but that the proofs related to astronomy. As an example, in one case, which had not been explained in previous astronomical and optical manuscripts, al-Qâyinî tried, despite lacking a vigorous proof, to show how a special point in a room could be lit up by sunlight throughout the year. His particular interest in astronomy led us to a general study of his other works, and it is worth noting that out of the nineteen works that have been attributed to him, eleven are devoted to astronomy and have not been thoroughly studied. Much more research is required on his astronomical manuscripts in order to obtain a better understanding of this author, the century in which he lived and the general state of science in the late Islamic period. After all, given the many as-yet unstudied manuscripts that still exist, maybe this decline was not as rapid as has previously been assumed.

## 1 INTRODUCTION

The common means of reconstructing the history of the Islamic sciences is mainly through Arabic references. A large portion of these were derived from Greek sources, especially from the eighth to the eleventh century A.D. (Jamili, 2006). But after this period there are some works in other languages too; particularly in Persian. So, for the study of the history of Islamic sciences in the latter period, research on these none-Arabic sources is also necessary. By such studies we can illustrate the history of many branches of science in the later Islamic eras (Kheirandish, 1998a: 125).

In Islamic optics we encounter a tradition of writing many books which have been titled *al-Manâẓir* and other similar titles that follow Euclid's geometric optics, which were translated from Greek during the Islamic period. There are currently a large number of these manuscripts. Kheirandish (ibid.) has investigated these, and her work is very important for the purposes of this paper.

Euclid's book on optics, *al-Manâẓir* (300 B.C.), is the first work that was translated into Arabic, and the tradition of writing books on this subject started through its influence (ibid.). These collections of successive manuscripts, that Kheirandish has titled 'Main texts', can be divided into three different types: (a) direct translations of Euclid's *al-Manâẓir*. For example, the first translation belongs to Hulail Ibn Sirjūn (A.D.827-900) and is titled *Fi Ikhtilaf al-Manâẓir* (Kheirandish, 1998a: 19); (b) related texts that besides the first kind form a tradition that from the tenth to the fourteenth centuries extended optics in the Islamic world. As an example, the work *Islah al-Manâẓir*, which is attributed to Abu Yūsuf al-Kindī (A.D.870), belongs to this group (Kheirandish, 1998b: 64); and (c) a collection of texts that related to Euclid's *al-Manâẓir* and covered different aspects of this field; through them this tradition and its historical and geographical borders will be known better. *Tahrir al-Manâẓir* of al-Ṭūsī (A.D.1273) and *Tajrīd al-Manâẓir* of Ibn Jarradeh (A.D.1277) are examples of this group (Kheirandish, 1998a: 19).

Besides these main texts, there are a number of manuscripts that complement texts in which some knowledge of mirrors and measurement are seen. These texts, which span a period of almost one thousand years and are found from Baghdad to Calcutta (presently Kallkata), date from a little before the first Arabic translations of Euclid's *al-Manâẓir* (Kheirandish, 1998a).

It is often considered that science stagnated in the Islamic world in the period after the fifteenth century A.D., when we cannot find any significant or influential works on optics or any other scientific fields (Dallal, 1999). The key question for us is: how was this tradition followed in this latter Islamic period, and which of the surviving manuscripts from this period provide the most prominent evidence of the tradition set by Euclid? Understanding the true nature and extent of this 'stagnation' is important.

To find the answer to this question an optical manuscript which belongs to these centuries was found (see Figure 1), and this has been edited and critically analyzed. According to previous studies, it could increase our historical knowledge of the progress of optics in the late Islamic period. It could also include some different aspects of theoretical and practical astronomy. Previous studies showed that a qualitative and quantitative review of its contents could be enough to show the continuation of the optical tradition in the seventeenth century set by Alhazen six century earlier (Kheirandish, 1998a: 51; 1998b: 133). This result encouraged us to continue Kheirandish's attempts to find the answer through a detailed study and editing of this manuscript. In addition, on the first page of this manuscript it is written that *Manâẓi*r *w*a *Marâyâ* is an important Persian treatise in this field with a lot of geometrical figures dated to A.D.1685 between pages 149 to 234, and that it was written at the Madreseh Charbagh in Mashhad.

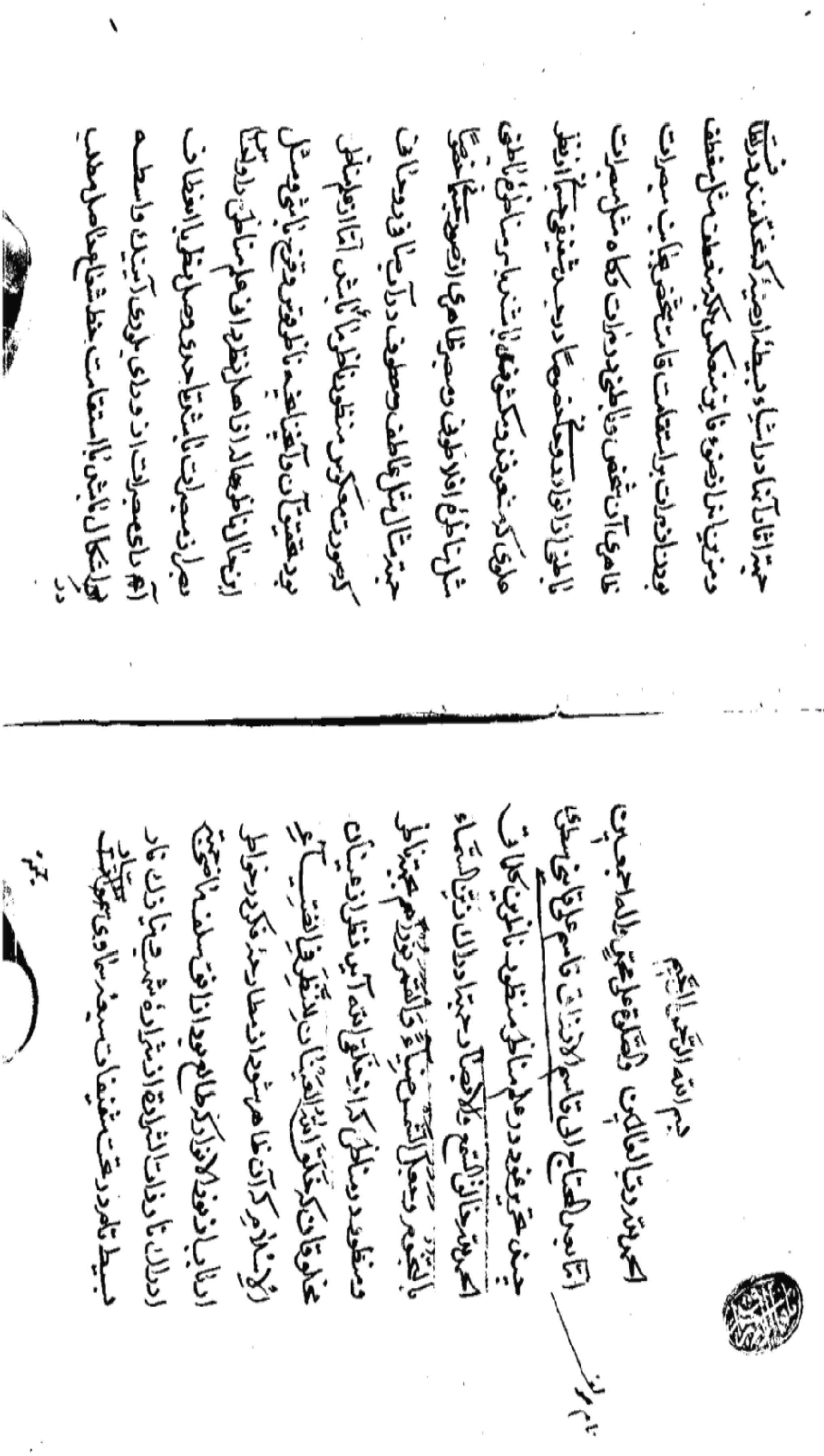

Figure 1: Pages 1 and 2 of the Qâsim cAlî al–Qâyinî's *Manâẓir wa Marâyâ* (courtesy: Central Library of the University of Tehran).

This treatise is now located in a collection of manuscripts with five others that will be mentioned later. Al-Qâyinî's seal appears on the first page and on page 149 in this collection. On the last page, these words have been written: "… this treatise was finished in the Madreseh Charbagh in Mashhad in Safar 17th in A.D.1685" Thus, we know the date when and the place where it was written. This manuscript, which consists of 85 pages each measuring 9 × 17 centimeters, has not been dedicated to any person, and the aim of its writing is not clear. The collection of which it forms a part is No.3530 in the Central Library at the University of Tehran and it originally belonged to Mr Bastani Rad who donated it to the Library. In the preface of the treatise it is noted that "From the evidence, this manuscript is the original and other copies of it have not been seen". Searches of different lists of manuscripts failed to reveal any other copies, so from current evidence we can conclude that this manuscript is unique.

The exact dates of al-Qâyinî's birth and death are not clear and there is not much information about him, although the bibliography and manuscript references do provide a little about different aspects of his life. Al-Qâyinî is known to have been a student of the fairly well-known astronomer and mathematician Muhammad Bagher ibn zany al-Abidin al-yazdi (d. A.D.1637) (Ansari and Ghori, 1987: 217), and he could not have died before A.D.1656 because in that year he wrote on the margin of the *Tashrīh dar Pargār*. It has been claimed by Rehatsek (1873, No. 21, p: 14-15) that he probably was born 18 years before A.D.1591, but this claim needs further investigation (see Storey, 1972: 90). From the available information about the places where he wrote his manuscripts mentioned on the first or last pages, it seems that he spent many years at the Madreseh Charbagh in Mashhad. In the seventeenth century A.D. this was an important school, and famous scholars taught there. In fact, during this century, the schools at Mashhad were the most important at that time. So, we can see that al-Qâyinî was living and studying at a time when the Mashhad schools were flourishing (Sayyed-kebari, 1999).

## 2 AL-QÂYINÎ'S SCIENTIFIC WORKS

### 2.1 Astronomical Works

2.1.1 *Muṭāliᶜ al-Muḍḥaj*: This is a Persian-language treatise on astronomy that dates to A.D.1677. There is a copy of this treatise in the Goharshad Library in Mashhad (Fazel, 1984-1992: 1146).

2.1.2 *Jāmiᶜ al-Anwār min al-Kawkab wa al-Abṣār*: This is a Persian treatise that Rehatsek (1873, No. 21, p: 14) has dated to A.D.1591. This is a treatise about observational tools, and mentions geometrical and mathematical arguments that were written in the year A.D.1686 (Storey, 1972: 90). Elsewhere this treatise is called the *Jāmiᶜ al-Anwār Ālāt-i Asṭurlāb*. There is a version of this treatise (No. 6460) in the Central Library of the Āstān Quds Raẓavī Library, in Mashhad (Osat Nateqi, 2006: 198), and there is another copy (No. 1-21) in the Mulla Firoze Collection in the K.R. Cama Oriental Research Library in Bombay, India (Ansari and Ghori, 1987: 217).

2.1.3 *Risālah dar ᶜIlm-i Hay'at*: There is a copy of this treatise (No.402) at the University of St. Petersburg in Leningrad (Storey, 1972: 90).

2.1.4 *Risālah dar Babi Isti^c^mal-i Asṭurlāb*: There is a copy of this treatise (No. 465/4) in the Tashkent Llibrary in Uzbekistan; another copy (No. 406/4) in the Central Library of the University of Tehran (ibid.); and a third copy (No. 669/1) in the Sipahsālār Library in Tehran (Munzavi, 1969: 230).

2.1.5 *Risālah dar Asṭurlāb*: There is a copy of this treatise (No. 403) at the University of St. Petersburg in Leningrad (Storey,1972: 90).

2.1.6 *Imtiḥān*: There is a copy of this treatise (No. 403) in St. Petersburg (ibid.). It is possible that this is the manuscript in which al-Qâyinî stated that he learnt the art of astrolabe-making from al-Yazdi, which is registered as No. PNS-114 in the State Public Library in Leningrad and has the title *Imtiḥān Asṭurlāb* (Ansari and Ghori, 1987: 217 and 223). As we can see the manuscripts number 2.1.5 and 2.1.6 have the same number; No. 403. It seems these two manuscripts belong to the collection which is numbered 403.

2.1.7 *Maṭlac-i Hīlāj*: This is a treatise about the astrolabe, and was written in A.D.1686. There is a version of this treatise (No. 2377/3) in the Majlis Library in Tehran (List of Manuscripts of the Majlis Library, 1970: 238).

2.1.8 *Risālah dar Mcarifat Qibla*: This treatise was copied in A.D.1647. There is a copy (No. 2377/2) of this treatise in the Majlis Library in Tehran (Munzavi, 1969: 331).

2.1.9 *Macrifat-i Taqvīm*: (al-Dharīac, 1983: 270).

2.1.10 *Matālīc al-Hikam*: This is a treatise about determining the times of prayers by using a sheet that al-Qâyinî has made by himself. There is also a copy of this treatise (No.6266) in the collection of the Majlis Lbrary in Tehran (Osat Nateqi, 2006: 198).

## 2.2 Mathematical Works

2.2.1 *Dhubdat al-Natāīj dar cIlm-i al-Hisāb*: This is a Persian treatise, and there is a version (No.F2) in the National Library of Fārs (Behrozi and Faghiri, 1972: 263).

2.2.2 *Tarjuma Wa Sharḥ-i al-Jabr Wa al-Muqābla Khwājā-i Naṣīr al-Din al-Ṭūsī*: There is a copy of this treatise (No.1319/2) in the Central Library at the University of Tehran which was written in A.D.1686. (Munzavi, 1969: 148).

2.2.3 *Tashrīh al-cAmāl*: In the list of manuscripts in the Āstān Quds Raẓavī Library this treatise is listed as *Tashriḥ dar Pargār* and compiled by Ghyāth al-Din Jamshīd Kāshānī or cAbdul-cAlī Bīrjandī, but it is ultimately mentioned that neither of them was probably the author. The subject of this treatise is explaining the *Pargār Mutanāseba*, which Mohammad Zaman Mashhadi wrote, with al-Qâyinî adding some marginal notes. There is a copy of this treatise (No.5266) in the Central Library of the Āstān Quds Raẓavī in Mashhad (Asef Fekrat, 1934, Volume 3: 312).

### 2.3 Other Works

2.3.1 *Khulāṣah al-Ikhlāṣ Fī Tafsīr Sūrah al-Ikhlāṣ*: This is a Persian treatise that was written in A.D.1686. There is a copy of this treatise (No. 6459) in the Central Library of the Āstān Quds Raẓavī in Mashhad (Asef Fekrat, 1934, Volume 1: 127).

2.3.2- *Manâẓir wa Marâyâ*: This is the Persian manuscript on optics that was written in A.D.1685 in Mashhad and is described in this paper (Daneshpajoh, 1961: 2549).

Besides the *Manâẓir wa Marâyâ*, there are four other manuscripts that belong to al-Qâyinî and comprise the collection of which *Manâẓir wa Marâyâ* is a part. Their names and the numbers in their first pages (in the same collection) are:

2.3.3 *Risālah fī Hudūs al-Asmā'*: (pp. 1-96 of Collection No. 3530) (ibid.).

2.3.4 *Dhakhīrah*: (pp. 100-107 of Collection No. 3530) (Daneshpajoh, 1961: 2550).

2.3.5 *Risālah dar cIlm-i al-Hurūf*: (pp. 107-118 of Collection No. 3530) (ibid.).

2.3.6 *Risālah Jafr Ṣaghīr wa Kabīr*: (pp. 119-148 of Collection No. 3530) (Daneshpajoh, 1961: 2551).

In a previous study on al-Qâyinî's works, thirteen manuscripts attributed to him were introduced (Ansari and Ghori, 1987). In this paper, six further manuscripts are identified.

## 3 A BRIEF ACCOUNT OF AL-QÂYINÎ WORKS ON OPTICS

### 3.1 Al-Qâyinî's Predecessors

In his book *De Aspectibus*, Abu Yusūf al-Kindī (ca. A.D. 873) proved Euclid's principles on optics and investigated the reflection of light rays by all kinds of mirrors. After him, all the famous scholars in this field emphasized the use of burning mirrors[1] (e.g. see Rashed, 1993). Ibn Sahl (A.D. 932 – 992) added burning instruments[1] to the burning mirrors. Ibn Sahl discovered that every material has a special optical property that is constant, namely the refractive index. Refraction of light was investigated after him by Alhazen (ibid.). Through his prominent book of optics, titled *al-Manâẓir*, the eleventh century A.D. scholar Alhazen is recognized as the most influential scholar in medieval optics. In medieval and early Europe, *al-Manâẓir* was very well received through Latin translations and it influenced the contents of other works on optics. Yet this influential book was largely unknown to later Muslims scholars, until the fourteenth century A.D. when Kamāl al-Dīn Fārsī (ca.A.D.1320) wrote his *Tanqīh al-Manâẓir* (see Lindberg, 1998). Quṭb al-Dīn Shīrāzī was Kamāl al-Dīn Fārsī's teacher, and in answering Fārsī's questions about optics introduced him to Alhazen's *al-Manâẓir*. Fārsī added three new chapters to this book and named his book *Tanqīh al-Manâẓir*. In this work, Fārsī investigated all the optical topics known at that time. This book and *al-Manâẓir* are the two best-known books on optics dating to the Islamic era, and published copies in Arabic are still available (Jabbar, 2005).

### 3.2 Al-Qâyinî's Optics: The *Manâẓir wa Marâyâ* Manuscript

*Manâẓir wa Marâyâ* is a Persian manuscript by al-Qâyinî that contains 42 different optical propositions or subjects, including reflection, refraction, shadows, the formation of rainbows and haloes, the dark room and the visual field of flat mirrors. This optical work drew on Alhazen's theories as outlined in *al-Manâẓir*. However, some new original material is included, such as the illumination of a room by sunlight during the year.

Al-Qâyinî's manuscript is divided into sixteen separate sections, each of which contains several similar topics and examples. The first six subjects note the basic principles and concepts of Greek and early Islamic optics. The next sections make use of these basics principles in describing experiments and natural phenomena. As we will see below, in fourteen of these experiments and explanations al-Qâyinî uses astronomical examples to introduce or explain optical concepts.

### 3.3 Optical-Astronomical Subjects in the *Manâẓir wa Marâyâ*

Those cases in which al-Qâyinî uses astronomical concepts and examples to explain optical propositions are listed below:

(1) Light from the Sun and Moon and meteors on terrestrial objects; haloes and rainbows as examples of optical cases.
(2) The existence of changing shadows on the Earth and the Moon, which are formed according to their positions relative to the Sun.
(3) Explaining how refraction of light makes the Sun appears to be larger soon after sunrise compared to later in the day when it is high in the sky.
(4) In reviewing the intensity and faintness of shadows, topics discussed include decreasing and increasing the amount of sunlight on the ground, moving shadows, shadow indices, using shadows to identify time (e.g. sun clocks) and how shadows on the Moon can be used as a scale by astronomers.
(5) The effects of the density of layers of air in producing different shades of sunlight in the twilight near the horizon, and apparent differences in the sizes of stars with elevation.
(6) Use of a dark room to study full and partial eclipses.
(7) Attempting to obtain the relationship between the angle of incidence and the angle of refraction at the surface of the water for a solar ray located at different heights above the surface of the water.
(8) A vague explanation of a strange optical-physical phenomenon: he explains that light rays which gather into a spherical shape make mass on the Moon!
(9) Explanation of the 'Moon disappearing' effect; like No.8, this phenomenon is very vague and physically unacceptable; he explains that when the light of the Moon is reflected from the surface of water, the Moon disappears, as seen from an observer close to the water. Explanation of such a vague effect has not been observed before, and its origin is uncertain.

(10) The effect of an observer's position relative to an object. Here al-Qâyinî is using the definition of parallax, and different types of it (see Kennedy, 1956).

(11) Explaining the curvature of the rainbow, based on particles between the sun and the observer, a mountain and the altitude of the sun.

(12) The formation of a halo around the Moon.

(13) Reviewing the changing angle of sunlight upon the surface of the Earth, due to the Sun's movement across the sky, and its effect on the production of shadows. This is similar to the method outlined in the *Surya-Siddhanta* for measuring the hypotenuse of an arc by using shadows in a dial (Burgess, 1858, 241).

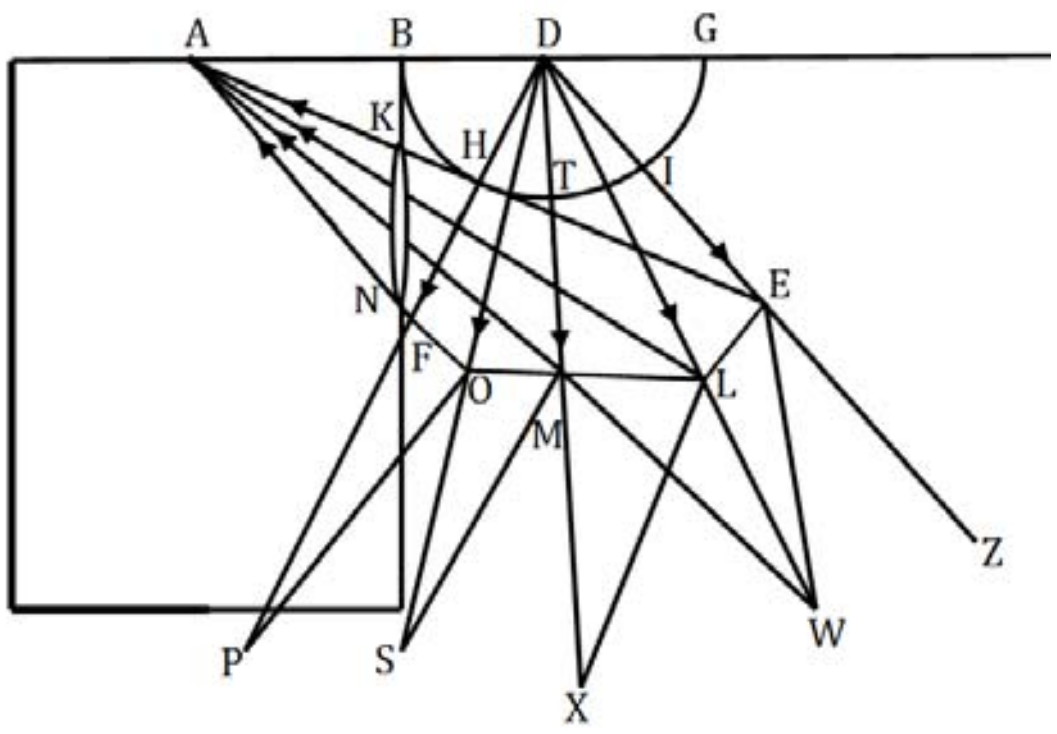

Figure 2: Al-Qâyinī's diagram to illustrate case number 14.

(14) Al-Qâyinî describes a geometrical method which focuses on using a particular combination of flat mirrors in front of the Sun to lead sunlight during the year from a hole in a wall into a room. From al-Qâyinî's explanation, it is not easy to understand how this works. It seems that it is an effort to illuminate a room but its actual use and the geometrical proof he provides must be questioned. The geometric description that he supplies (see Figure 2) is very vague and contains some errors. For example, the exact place of the hole in the wall or the ceiling is not clearly specified. Referring to Figure 2, AG is a line parallel to the meridian of the location. Point D is the position along this line where the Sun's rays enter. Rays from the Sun at one solstice at sunrise or sunset are DE and DF, and are reflected from mirrors EL and FO and then pass through the hole KN to reach A. The other rays follow similar paths to point A. As is obvious, there are lots of geometrical, optical and astronomical errors in this figure. For example, the equality of the angle of incidence angle and the angle of reflection is not considered.

There are several other mistakes in *Manâẓir wa Marâyâ*. For example, in the initial definition of the refractive angle from a surface, al-Qâyinî considers it equal to the angle of reflection, but in the other parts of his manuscript he does not use this definition but instead adopts Alhazen's method. Geometrical mistakes also occur in proving some propositions. For instance, when he wants to show why an object is seen bigger than the original, in describing how a burning instrument is made from flat mirrors, al-Qâyinî has a significant mistake in his proof. The vaguest part of his manuscript relates to the geometrical proof of the illuminated room, which is supposedly explained in Figure 2.

## 4 DISCUSSION AND CONCLUDING REMARKS

In his manuscript, al-Qâyinî mentions the names of some of his predecessors, who may have influenced his knowledge of optics. For example, he cites Alhazen as the most important scholar in this field, but apart from him he also refers to Euclid, Galen and al-Kindī.

Among the 42 optical propositions considered in his manuscript al-Qâyinî used fourteen astronomical and meteorological propositions, but these are often too general and usually are explained using examples rather than rigorous logical proofs. All of the explanations are qualitative and in some cases only a simple geometric proof is presented. The diversity and order of subjects seems to have been irrelevant. Perhaps this is due to the fact—contrary to previous belief—that al-Qâyinî was not an optician, and that he was merely trying to answer questions that personally interested him.

It is interesting that *Manâẓir wa Marâyâ* is not dedicated to anyone and thus far only one copy of it has been found, thereby strengthen the idea that al-Qâyinî should not be identified as an optical researcher (Akbari, 2008). We were therefore persuaded to study his other works in order to help us understand his reason for writing *Manâẓir wa Marâyâ*, and so that we might obtain a better viewpoint on his profession.

Upon reviewing *Manâẓir wa Marâyâ* alongside the other works, it is found that of the nineteen other treatises that are known, eleven deal with the natural sciences (astronomy and optics) and three with arithmetic, while the remainder are not thematically linked but deal with such diverse topics as the esoteric sciences and interpretation of the Quran.

Furthermore, of the eleven treatises that relate to the natural sciences, ten are about astronomy and only the one deals with optics. This clearly indicates that al-Qâyinî's main scientific interest was astronomy. And as we saw above, in fourteen instances he used astronomical examples and meteorological phenomena to explain optical theorems. In all, 35% of all definitions and theorems in this manuscript relate to optical-astronomical topics.

Examination of the dates of al-Qâyinî's astronomical treatises shows that they were written throughout his life, and this observation, plus the fact that more than half of his works were about astronomy, would seem to indicate that he should be regarded as an astronomer (especially since three other manuscripts deal with arithmetic and one with optics).

There are a lot of unanalyzed details of al-Qâyinî's works and his place in the seventeenth century A.D. Islamic science, and through further studies we hope to find out more about his scientific personality and the situation of the sciences at this time. In order to make a more accurate assessment of his scientific efforts it is necessary to study all of his astronomical treatises. Such a study should reveal whether the other treatises were also written generally, and with incomplete proofs, or whether he really was an accurate astronomer. If al-Qâyinî was a careful astronomer, then his writing on other topics was simply out of personal curiosity; if he was not, then he can best be described as an imprecise combiner of his predecessors' works.

Evaluating the works of Islamic scientists and their positions in the field of science is very difficult because we have very little knowledge of their manuscripts. There are a large volume of resources and manuscripts that have yet to be accurately reviewed. For example, if we consider al-Qâyinî to be a seventeenth century scholar, it is telling that only three of his manuscripts have been studied: a work on astronomical instruments (Ansari and Ghori, 1987), an interpretation of a part of the Quran (Nateqi, 2006) and his *Manâẓir wa Marâyâ*. Clearly he dedicated most of his time to astronomy, and as such probably he considered himself to be an astronomer. But in order to provide a realistic interpretation of our late Islamic scientific legacy (i.e. from the fifteenth to the nineteenth century A.D.), it is important that specialists in manuscripts, historians and particularly historians of science do all in their power to find, list and research manuscripts that date to this interval.

Another question can perhaps also be answered in the process and that is whether there really was a decline in Islamic science at this time. In other words, was this a period of stagnation as some historians of science have claimed (e.g. see Dallal, 1999), or were innovations still occurring? The extent to which there was stagnation in astronomy has been queried by some historians of astronomy in recent years, and they argue that from the sixteenth century A.D. onwards Muslim astronomers made some advances (Ragep, 2001; Saliba, 2000), but that the emphasis shifted from theoretical aspects to observational astronomy.

One way to test the validity of these claims would be to review the surviving astronomical literature and resources from this period. Unfortunately, much of this material remains unknown and untouched. Some of it is in personal libraries where it is unlisted, inaccessible and completely unknown to historians of science.

## 5 ACKNOWLEDGEMENT

This research was carried out as part of my Master's Thesis at the University of Tehran. I am very grateful to my supervisor, Iraj Nikseresht and my consulting supervisors, Gholamreza Jamshid-Nejad Avval and Hamid-Reza Giahi Yazdi, whose encouragement, guidance and support from beginning to end enabled me to develop a deep understanding of research in the history of science. I also wish to express my deepest gratitude to S.M. Razaullah Ansari and Mohammad Bagheri for their kind guidance, advice and support before the conference, and for helping me complete and edit this paper. I also wish to thank Mohammad Akhlaghi, who offered his support in a number of ways at all levels of my research.

## 6 NOTES

1. A burning mirror is a flat, concave or convex mirror that can concentrate the sun's rays onto a small area by using reflecting surfaces to focus the light, heating up the area and thus resulting in ignition of the exposed surface. Burning instruments; burning glass, burning lens, and burning mirrors are examples for this kind of instruments.